\pdfoutput=1
\documentclass[12pt]{iopart}
\usepackage{graphicx,dcolumn,iopams,setstack,amssymb,amsfonts,epsfig,float,subfigure}
\usepackage{amsthm}
\usepackage{txfonts}
\usepackage[T1]{fontenc}
\usepackage[english]{babel}
\usepackage{bm}

\begin{document}
\title[Invariant difference schemes]{Invariant difference schemes and their application to $SL(2,\mathbb{R})$ invariant ordinary differential equations}
\author{R REBELO$^{a,b}$ and P WINTERNITZ$^{a,b}$}
\address{$^a$ Département de mathématiques et de statistique, Université de Montréal,
 C.P. 6128, succ. Centre-ville, Montréal, Québec H3C 3J7, Canada.}
\address{$^b$ Centre de recherches mathématiques, Université de Montréal,
 C.P. 6128, succ. Centre-ville, Montréal, Québec H3C 3J7, Canada.}
\eads{\mailto{raph.rebelo@gmail.com}, \mailto{wintern@crm.umontreal.ca}}

\begin{abstract}
We present an exposition of a method of discretizing ordinary
differential equations while preserving their Lie point symmetries.
This method is very general and can be applied to any ODE with a
nontrivial symmetry group. The method is applied to obtain numerical
slutions of second and third order ODEs invariant under two different
realizations of $SL(2,\mathbb{R})$. The symmetry preserving method is shown to
provide a better qualitative description of solutions than standard
methods. In particular it provides solutions that are valid close to
singularities and beyond them.
\end{abstract}

\submitto{Journal of Physics A: Mathematical and Theoretical}
\maketitle

\newpage
\section{Introduction}

Lie group theory was originally invented as a systematic tool for obtaining exact analytical solutions of ordinary and partial differential equations(ODEs and PDEs). For ODEs the existence of a nontrivial symmetry group (local Lie group of local point transformations taking solutions into solutions) makes it possible to reduce the order of the equation. If the symmetry group is large enough, the problem can be reduced to quadratures \cite{1}. The reduction to quadratures in principle provides the general solution of the ODE. This may however be in implicit form that is of little use in visualizing the solution, or presenting it in the form of a graph. What happens in such cases is basically that an ODE is transformed into an algebraic, or transcendental equation.\\

If the symmetry group is not large enough, or its structure is such that the solution provided by the group is too implicit to be useful, it is necessary to resort to numerical solutions. The question arises: can the symmetry group still be put to good use?\\

All numerical methods for solving ODEs replace the differential equation by a difference one, usually on an a priori chosen lattice, either a regular one, or one adapted to some known or expected behaviour of solutions. Most or all Lie point symmetries are lost in this procedure.\\

Over the last 20 years a considerable effort has been made to apply Lie group theory to difference equations (for reviews containing references to the original articles see \cite{2,3}). One approach to this problem is to generate {\it invariant difference schemes}, consisting of two equations, involving independent and dependent variables $(x_n,y_n)$ evaluated at $N+1$ points (in order to be able to approximate an ODE of order $N$)
\begin{eqnarray}\label{Ea}
E_a(x_{n+K},...,x_{n+L},y_{n+K},...,y_{n+L})=0, \quad a=1,2, \quad L-K=N-1.
\end{eqnarray}

The equations must allow to calculate $x_{N+L},y_{N+L}$ if all previous points are known. This is a condition on the Jacobian matrix, namely
\begin{eqnarray}
rank\left(\frac{\partial (E_1,E_2)}{\partial \left(x_{n+K},...,x_{n+L-1},y_{n+K},...,y_{n+L-1}\right) }\right)=2.
\end{eqnarray}

Thus the scheme \eref{Ea} represents a difference equation and an equation for the lattice.\\

Let us consider a Lie algebra $g$ realized by vector fields in two variables
\begin{eqnarray}\label{vector}
X_i=\xi_i(x,y)\partial_x+\phi_i(x,y)\partial_y \qquad i=1,...,M=dim(g)
\end{eqnarray}
where $x$ and $y$ are respectively the independent and dependent variables in a differential, or difference equation.\\

Invariant ODEs of order $N$ are obtained as follows. We prolong the vector field \eref{vector} to order $N$
\begin{eqnarray}\label{4}
pr^{N}X_i=X_i+\phi_i^x\partial_{y'}+\phi_i^{xx}\partial_{y''}+...+\phi_i^{Nx}\partial_{y^{(N)}},
\end{eqnarray}
where the coefficients $\phi^x,...,\phi^{Nx}$ are expressed in terms of partial derivatives of the coefficients $\xi,\phi$ in the original vector field \eref{vector} \cite{1}. Invariants $I_{\mu}(x,y,y_x,...,y_{Nx})$, $1\leq\mu\leq A$, of the group action corresponding to Lie algebra $g$ are obtained by solving the system of first order PDEs
\begin{eqnarray}\label{pr}
prX_i\phi(x,y,y_x,...,y_{Nx})=0, \qquad i=1,...,M.
\end{eqnarray}
The number of functionally independent solutions of \eref{pr} is at least $A=N+2-M$. It can be larger if the M equations \eref{pr} are linearly dependent on some manifold $S$ in the corresponding jet space. The invariant equation will have the form
\begin{eqnarray}\label{F}
F(I_1,...,I_A)=0
\end{eqnarray}
where $F$ is any sufficiently smooth function.\\

An invariant difference scheme is obtained in a similar manner. We write the vector fields \eref{vector} in one chosen point $n$ then prolong them to as many points as figure in the scheme (i.e. $N+1$). We have
\begin{eqnarray}\label{prd}
pr_D\ X_{i,n}=\sum_{k=K}^L\left\{\xi_i(x_{n+k},y_{n+k})\partial_{x_{n+k}}+\phi_i(x_{n+k},y_{n+k})\partial_{y_{n+k}}\right\} \qquad 1\leq i\leq M. 
\end{eqnarray}

The {\it discrete invariants} $I_{n,\alpha}(x_{n+j},y_{n+j})$, $1\leq \alpha \leq B$, $K\leq j\leq L$ are obtained by solving the system of equations
\begin{eqnarray}\label{prdis}
pr_D\ X_{i,n}\phi(x_{n+j},y_{n+j})=0.
\end{eqnarray}
The number of functionally independent solutions of \eref{prdis} is at least $B=2N+2-M$ and is larger if the equations \eref{prdis} are linearly dependent on some manifold $\tilde{S}$ in the discrete jet space.\\

In the continuous limit \eref{prd} reduces to \eref{4} so the discrete invariants will reduce to the continuous ones. In general there are more discrete invariants than continous ones, so some combinations of the discrete invariants will go to zero in the limit, others will go to the continuous invariants. This makes it possible to choose an appropriate basis for the discrete invariants and to write an invariant difference scheme:
\begin{eqnarray}
E_1=F(I_{n,1},...,I_{n,A})=0\\
E_2=E_2(I_{n,1},...,I_{n,A})=0
\end{eqnarray}
with $F$ as in \eref{F} and $E_2$ satisfying $E_2\rightarrow 0$ in the continuous limit.\\

Such difference systems may be of interest in their own right and describe discrete phenomena on some specific symmetry adapted lattice. On the other hand the difference scheme may be chosen to have a specific ODE as its continuous limit. By construction the ODE and the difference scheme will be invariant under the same symmetry group $G$. Solving the difference scheme numerically provides approximate numerical solutions of the ODE. Since the symmetry group $G$ determines many properties of the solution space one can expect that numerical schemes using a symmetry adapted discretization will have some advantages over other numerical methods. It has indeed been shown that for first order ODEs symmetry preserving discretizations are exact: the invariant differential equations and difference schemes have exactly the same solutions \cite{4}. Symmetry preserving discretizations of second order ODEs can be solved exactly using a Lagrangian approach \cite{5,6}. These analytic solutions of the difference schemes then converge rapidly to the solutions of the ODEs \cite{6}. Two recent articles \cite{7,8} were devoted to numerical solutions of second and third order ODEs. It was shown (at least for the considered examples) that the qualitative behavior of solutions of the ODEs, specially in the neighbourhood of singularities, is better described by symmetry preserving discretizations than by standard methods.\\

Four inequivalent realizations of $sl(2,\mathbb{R})$ by vector fields of the form \eref{vector} exist \cite{9}. In this paper we concentrate on two of them, not treated in previous articles \cite{7,8}. We construct their differential invariants up to order three and their difference invariants involving up to four points. This allows us to write all invariant ODEs of order up to three and their discretizations.\\

In Section 2 we present the four realizations of $sl(2,\mathbb{R})$. The invariant ODEs and their discretizations are presented in Section 3. Section 4 is devoted to numerical solutions and Section 5 to conclusions.\\

\section{The four realizations of sl(2,$\mathbb{R}$)}

Let $\{X_1,X_2,X_3\}$ be three vector fields of the form \eref{vector} satisfying the commutation relations
\begin{eqnarray}
[X_1,X_2]=X_1, \quad [X_2,X_3]=X_3, \quad [X_1,X_3]=2X_2. 
\end{eqnarray}
One of them, say $X_1$ can be straightened out to $X_1=\partial_y$. Then $X_2$ can be transformed either into $X_2=y\partial_y$ or $X_2=x\partial_x+y\partial_y$. Point transformations leaving the standardized fields $X_1$ and $X_2$ invariant will further simplify $X_3$ and we obtain the four inequivalent realizations, namely:\\

1. $sl_1(2,\mathbb{R})$:
\begin{eqnarray}\label{2.2}
X_1=\partial_y, \quad X_2=y\partial_y, \quad X_3=y^2\partial_y.
\end{eqnarray}
The three vector fields \eref{2.2} are linearly connected, that is in any given point of $\mathbb{R}^2$ they are linearly dependent. This $sl(2,\mathbb{R})$ algebra is not maximal among finite dimensional subalgebras of $diff(2,\mathbb{R})$ but can be imbedded into $sl_x(2,\mathbb{R})\oplus sl_y(2,\mathbb{R})$ with
\begin{eqnarray}
sl_x(2,\mathbb{R})=\{\partial_x,x\partial_x,x^2\partial_x\}.
\end{eqnarray}
For the remaining three $sl(2,\mathbb{R})$ algebras no two of the three vector fields $X_1$, $X_2$ and $X_3$ are linearly connected.\\

2. $sl_2(2,\mathbb{R})$:
\begin{eqnarray}
X_1=\partial_y, \quad X_2=x\partial_x+y\partial_y, \quad X_3=2xy\partial_x+y^2\partial_y.
\end{eqnarray}
This $sl(2,\mathbb{R})$ algebra is not maximal in $diff(2,\mathbb{R})$. We can add $X_4=x\partial_x$ and obtain the algebra $gl(2,\mathbb{R})$. The algebra is imprimitive in that the coefficients of $\partial_y$ are all functions of $y$ alone. The corresponding $SL(2,\mathbb{R})$ group action allows an invariant foliation.\\

3. $sl_3(2,\mathbb{R})$:
\begin{eqnarray}
X_1=\partial_y, \quad X_2=x\partial_x+y\partial_y, \quad X_3=2xy\partial_x+(-x^2+y^2)\partial_y.
\end{eqnarray}

4. $sl_4(2,\mathbb{R})$:
\begin{eqnarray}
X_1=\partial_y, \quad X_2=x\partial_x+y\partial_y, \quad X_3=2xy\partial_x+(x^2+y^2)\partial_y.
\end{eqnarray}
These two realizations are equivalent over $\mathbb{C}$ but not over $\mathbb{R}$. They are both primitive and both are maximal subalgebras of $diff(2,\mathbb{R})$.\\

The ODEs invariant under $SL_1(2,\mathbb{R})$ and $SL_2(2,\mathbb{R})$ were treated earlier \cite{7,8}. Here we concentrate on $SL_3(2,\mathbb{R})$ and $SL_4(2,\mathbb{R})$.

\section{Invariant ODEs and difference schemes}  

\subsection{$sl_3(2,\mathbb{R})$: $X_1=\partial_y$, $X_2=x\partial_x+y\partial_y$, $X_3=2xy\partial_x+(y^2-x^2)\partial_y$}

A complete set of functionally independent differential invariants up to third order is
\begin{eqnarray}\label{sl3inv}
I_1=\frac{y'(1+y'^2)-xy''}{(1+y'^2)^{3/2}}, \qquad I_2=\frac{3x^2y'y''^2-x^2y'''(1+y'^2)}{(1+y'^2)^3}.
\end{eqnarray}

Note that a complete family up to any order can then be deduced using P. Olver's Proposition 2.53 stated in Section 2.5 of \cite{1}. This is true for all realizations.\\

In the discrete case a basis for all 3 point invariants is
\begin{eqnarray}\label{sl3disinv}
I_1^n=\left(\frac{(x_n-x_{n-1})^2+(y_n-y_{n-1})^2}{x_nx_{n-1}}\right)^{1/2}, \qquad I_1^{n+1}=\left(\frac{(x_{n+1}-x_n)^2+(y_{n+1}-y_{n})^2}{x_{n+1}x_{n}}\right)^{1/2},\nonumber\\
I_2^{n+1}=\left(\frac{(x_{n+1}-x_{n-1})^2+(y_{n+1}-y_{n-1})^2}{x_{n+1}x_{n-1}}\right)^{1/2}.
\end{eqnarray}

A complete family for any number of points can then be obtained simply by shifting the above invariants. Namely, the shifts of $I_1^{n+1}$ and of $I_2^{n+1}$ would be the two new invariants for a 5 points scheme.\\

Combinations of the discrete invariants \eref{sl3disinv} that approximate $I_1$ and $I_2$ from \eref{sl3inv} are
\begin{eqnarray}
J_1^{n+1}&\equiv \left(-8\frac{I_2^{n+1}-(I_1^n+I_1^{n+1})}{I_1^nI_1^{n+1}(I_1^{n+1}+I_1^n)}+1\right)^{1/2}\quad \text{and}\\
J_2^{n+2}&\equiv \frac{3}{I_1^{n}+I_1^{n+1}+I_1^{n+2}}\left(J_1^{n+2}-J_1^{n+1}\right)\nonumber
\end{eqnarray}
respectively.\\

The invariant 2nd order ODE is
\begin{eqnarray}\label{sl32ndode}
I_1=\frac{y'(1+y'^2)-xy''}{(1+y'^2)^{3/2}}=C
\end{eqnarray}
where $C$ is an arbitrary constant. To get rid of possible sign ambiguities, we solve the square of \eref{sl32ndode} $I_1^2=C^2$ to obtain the solution
\begin{eqnarray}\label{sl22eordresol2}
(y-y_0)^2+(x\pm C/a)^2=1/a^2
\end{eqnarray}
with $a,y_0$ integration constants and $a\neq 0$. Those are circles with center $(\mp C/a,y_0)$ and radius $r=1/a$.\\

An O$\triangle$S that goes to the ODE \eref{sl32ndode} in the continous limit is obtained if we put
\begin{eqnarray}\label{a}
J_1^{n+1}\equiv\left(-8\frac{I_2^{n+1}-(I_1^n+I_1^{n+1})}{I_1^nI_1^{n+1}(I_1^{n+1}+I_1^n)}+1\right)^{1/2}=C,\quad
E_2(I_1^n,I_1^{n+1},I_2^{n+1})=0
\end{eqnarray}
with $E_2$ defining the mesh and going to $0$ in the continuous limit.\\

The 3rd order invariant ODE can be written as
\begin{eqnarray}\label{sl33rdode}
\frac{3x^2y'y''^2-x^2y'''(1+y'^2)}{(1+y'^2)^3}=F\left(\frac{y'(1+y'^2)-xy''}{(1+y'^2)^{3/2}}\right).
\end{eqnarray}

An O$\triangle$S that goes to the ODE \eref{sl33rdode} in the continous limit is obtained if we put
\begin{eqnarray}
J_2^{n+2}=\frac{3}{I_1^n+I_1^{n+1}+I_1^{n+2}}(J_1^{n+2}-J_1^{n+1})=F(J_1^{n+1})
\end{eqnarray}
where $J_1^{n+1}$ is given in \eref{a} and the lattice is
\begin{eqnarray}\label{b}
 E_2(I_1^n,I_1^{n+1},I_1^{n+2},I_2^{n+1},I_2^{n+2})=0
\end{eqnarray}
with $E_2$ going to $0$ in the continuous limit.

\subsection{$sl_4(2,\mathbb{R})$: $X_1=\partial_y$, $X_2=x\partial_x+y\partial_y$, $X_3=2xy\partial_x+(x^2+y^2)\partial_y$}

A complete set of functionally independent differential invariants up to third order is
\begin{eqnarray}\label{sl4inv}
I_1&=\frac{xy''+y'(y'^2-1)}{(y'^2-1)^{3/2}},\\
I_2&=\frac{2x^2(y'+1)y'''+3((y'-1)(y'+1)^2(3y'^2-1)+4xy'(y'+1)y''-2x^2y''^2)}{(y'-1)^2(y'+1)^3}\nonumber
\end{eqnarray}
and the second and third order ODEs are
\begin{eqnarray}
I_1=C,\label{sl42ndode}\\
I_2=F(I_1).\label{sl43rdode}
\end{eqnarray}

We again take the square of the second order equation $I_1^2=C^2$ and obtain the solution
\begin{eqnarray}\label{sl32eordreodesol}
(x\pm C/a)^2-(y-y_0)^2=1/a^2
\end{eqnarray}
with $a\neq 0,y_0$ integration constants. The solutions for $a\neq 0$ are hyperbolas.\\

In the discrete case a complete set on 3 points is given by
\begin{eqnarray}\label{sl4disinv}
I_1^n=\left(\frac{(y_n-y_{n-1})^2-(x_n-x_{n-1})^2}{4x_nx_{n-1}-((y_n-y_{n-1})^2-(x_n-x_{n-1})^2)}\right)^{1/2},\nonumber\\ I_1^{n+1}=\left(\frac{(y_{n+1}-y_{n})^2-(x_{n+1}-x_{n})^2}{4x_{n+1}x_{n}-((y_{n+1}-y_{n})^2-(x_{n+1}-x_{n})^2)}\right)^{1/2},\\
I_2^{n+1}=\left(\frac{(y_{n+1}-y_{n-1})^2-(x_{n+1}-x_{n-1})^2}{4x_{n+1}x_{n-1}-((y_{n+1}-y_{n-1})^2-(x_{n+1}-x_{n-1})^2)}\right)^{1/2}.\nonumber
\end{eqnarray}

Combinations of the discrete invariants \eref{sl4disinv} that approximate $I_1$ and $I_2$ from \eref{sl4inv} are
\begin{eqnarray}
J_1^{n+1}&\equiv\sqrt{2}\left(\frac{I_2-(I_1+I_{1+})}{I_1I_{1+}(I_1+I_{1+})}-1\right)^{1/2},\\
J_2^{n+2}&\equiv\frac{3}{I_1^n+I_1^{n+1}+I_1^{n+2}}(J_1^{n+2}-J_1^{n+1})+6J_1^2+3\nonumber
\end{eqnarray}
so the corresponding O$\triangle$S are respectively
\begin{eqnarray}
J_1^{n+1}=C, \quad J_2^{n+2}=F(J_1^{n+1})
\end{eqnarray}
where $C$ and $F$ are the same as in \eref{sl42ndode} and \eref{sl43rdode}. An invariant equation for the mesh must be added in each case as in \eref{a} and \eref{b}.

\section{Numerical solutions}

To perform numerical tests, the arbitrary function $F$ of Section 3 needs to be specified. We choose $F(I_1)=I_1^2$ (this choice is arbitrary).

The second and third order ODEs are then discretized in several different ways: standard finite difference methods, matlab solver (ode45) which uses Runge-Kutta methods of order four and five, and the symmetry preserving discretization which has been described in the previous sections. The standard finite difference method consists of approximating the derivatives of the dependant variable using Lagrange interpolation polynomials (see \cite{7} for a more detailed explanation of the numerical methods used). As a quick reminder, the standard finite difference method gives on a four point scheme
\begin{eqnarray}
y'(x_{n+1/2})\approx\frac{1}{24h}(27(y_{n+1}-y_n)-(y_{n+2}-y_{n-1})),\nonumber \\
y''(x_{n+1/2})\approx\frac{1}{2h^2}(y_{n+2}-(y_{n+1}+y_{n})+y_{n-1}),\\
y'''(x_{n+1/2})\approx\frac{1}{h^3}(y_{n+2}-3y_{n+1}+3y_{n}-y_{n-1}),\nonumber
\end{eqnarray}
where $x_{n+1/2}=\frac{x_n+x_{n+1}}{2}$ is the scheme's center.\\

We will be interested in the behaviour of solutions near singularities (blow up in the first derivative).\\

\subsection{$sl_3(2,\mathbb{R})$} 

Figure 1 shows the behaviour of the standard and symmetric methods for \eref{sl32ndode}. The choice for $a$ corresponds to some $y_0'$ since a is the integration constant. The exact solution is \eref{sl22eordresol2} and is represented by the continuous line. The Newton method applied to the standard scheme converges poorly (and even fails to converge for $h\sim 0.01$ and smaller). It stops to describe the behaviour of the solution correctly near the first derivative blow up. The symmetric method integrates around the circle without any difficulties.\\

The mesh equation for the symmetric method is chosen to be
\begin{eqnarray}\label{7}
E_2(I_1^n,I_1^{n+1},I_2^{n+1})\equiv I_1^{n+1}-I_1^{n}=0.
\end{eqnarray}
This mesh equation is also chosen for the 3rd order equation and for both equations invariant under $sl_4$.

Figure 2 shows the behaviour of each method for the 3rd order equation
\begin{eqnarray}\label{8}
x^2(1+y'^2)y'''-x^2(3y'-1)y''^2-2xy'(1+y'^2)y''+y'^2(1+y'^2)^2=0.
\end{eqnarray}

Matlab solver (ode45) encounters a singuarity near $x=1.28$. The solution stays finite, but, again, there is a blow up in it's first derivative.  While ode45 stops integrating and the standard method blows up, the symmetric method integrates through the singularity and stays finite.\\ 

\begin{figure}[htbp]
\begin{center}
\includegraphics[width=12cm,height=12cm]{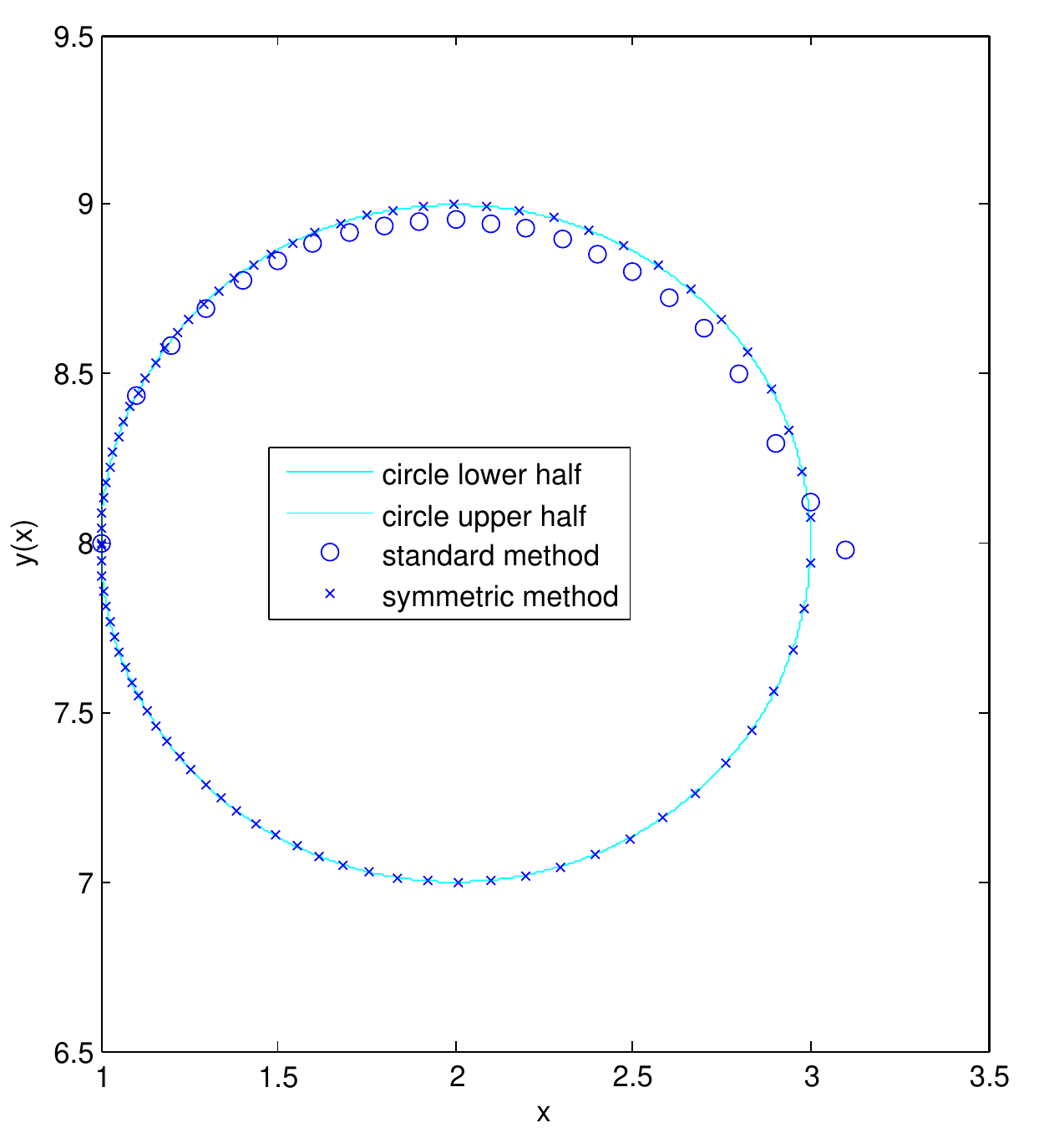}
\caption{Symmetric and standard method for \eref{sl32ndode} with initial conditions $\{x_0=1,y_0=8,C=2,a=1\}$}
\label{fig1}
\end{center}
\end{figure}

\begin{figure}[htbp]
\begin{center}
\includegraphics[width=12cm,height=12cm]{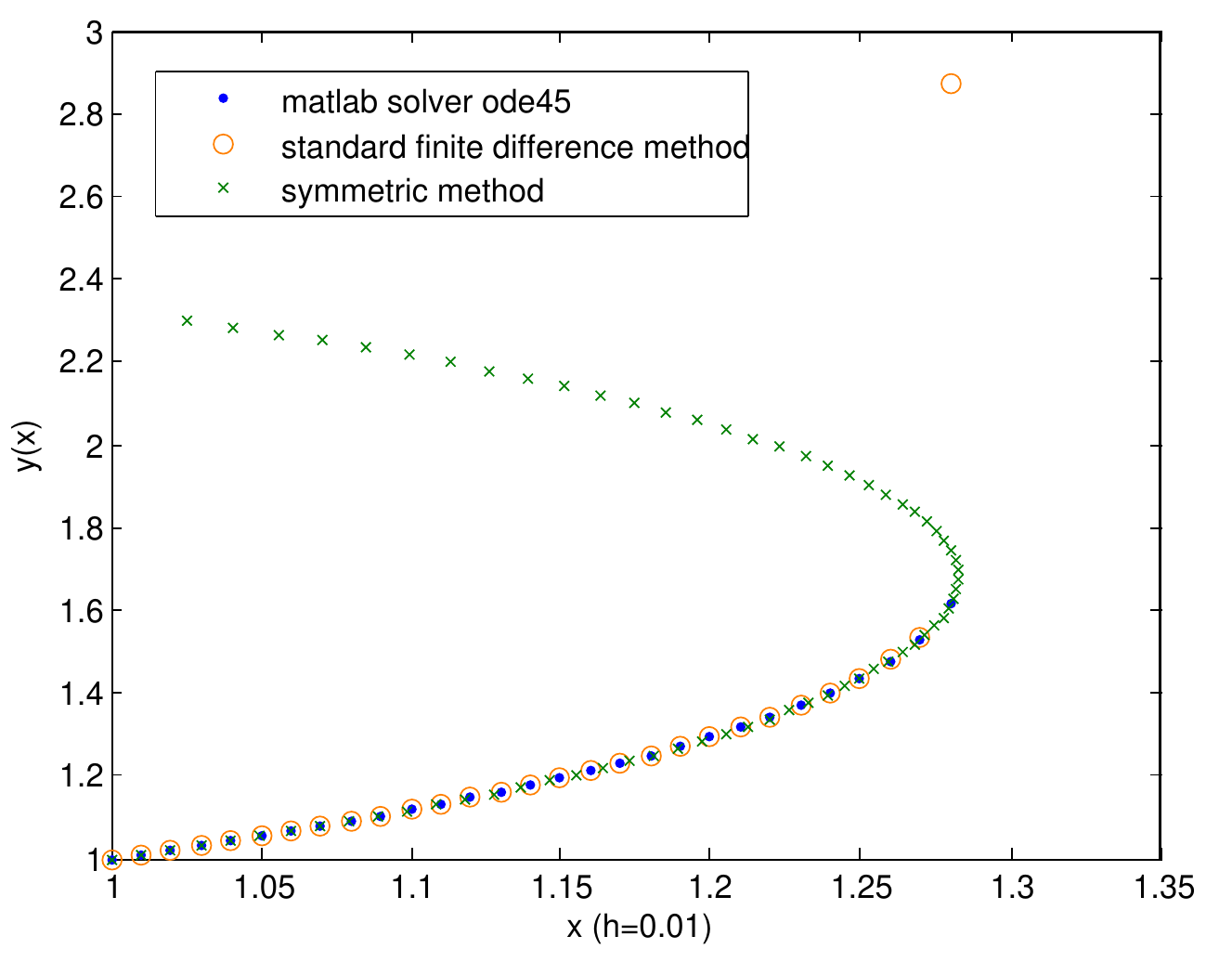}
\caption{All methods for \eref{8} with initial conditions $\{x_0=1,y_0=1,y'_0=1,y''_0=3\}$}
\label{fig2}
\end{center}
\end{figure}

While the first example is didactic since the analytic solution is known, the second one shows that the nice behaviour of the symmetric method holds for the more complex equation \eref{sl33rdode}.\\

Moreover, note the relative simplicity of the symmetric scheme for the 3rd order equation
\begin{eqnarray}\label{sl2syst}
(x_{n+1}(2+I_1^2)-x_{n}(2+\beta_n^2))x_{n+2}+2(y_{n+1}-y_{n})y_{n+2}=x_{n+1}^2-x_{n}^2+y_{n+1}^2-y_{n}^2\\
(x_{n+2}-(1+I_1^2/2)x_{n+1})^2+(y_{n+2}-y_{n+1})^2=(1+I_1^2/4)I_1^2x_{n+1}^2\nonumber
\end{eqnarray}
where $\beta_n$ and $I_1$ are some constants at each step. The only unknowns in this system are $x_{n+2}$ and $y_{n+2}$. Thus, solving the system \eref{sl2syst} amounts to finding the intersection between a straight line and a circle at each step while the standard finite difference scheme involves a Newton iteration on a line counting several hundred characters. Matlab solver, while very precise, also has a high computationnal cost. There are known numerical algorithms to find conic intersections (we used \cite{10}). The geometrical similarity between the exact solution for the 2nd order equation and this scheme is also interesting.\\

\subsection{$sl_4(2,\mathbb{R})$}

Figure 3 shows the behaviour of the standard and symmetric methods for \eref{sl42ndode}. The standard method stops correctly describing the solution near the blow up in the first derivative (it rapidly diverges after the strange behaviour shown in the figure). The symmetric method integrates on the entire branch of the hyperbola without any difficulties.\\

Figure 4 shows the behaviour of each method for the 3rd order equation
\begin{eqnarray}\label{9}
2x^2(y'^2-1)y'''+(y'^2-1)^2(8y'^2-3)+10xy'y''(y'^2-1)-x^2y''^2(6y'-5)\\
=0\nonumber
\end{eqnarray}

Again, the symmetric method integrates through the blow up in the first derivative while the other methods stop integrating.\\

\begin{figure}[htbp]
\begin{center}
\includegraphics[width=12cm,height=12cm]{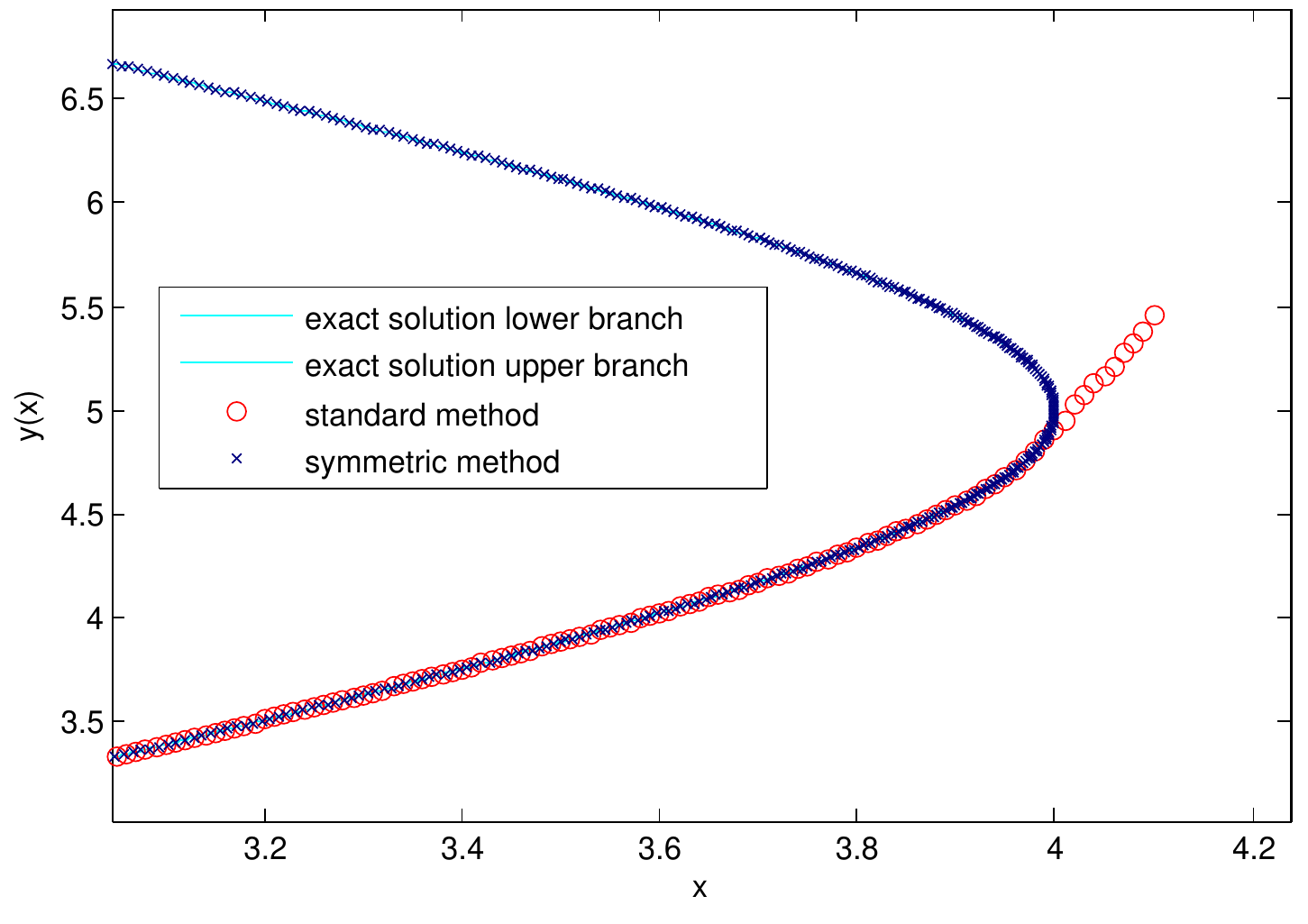}
\caption{Symmetric and standard method for \eref{sl42ndode} with initial conditions $\{x_0=2,y_0=5,C=5,a=1\}$}
\label{fig3}
\end{center}
\end{figure}

\begin{figure}[htbp]
\begin{center}
\includegraphics[width=12cm,height=12cm]{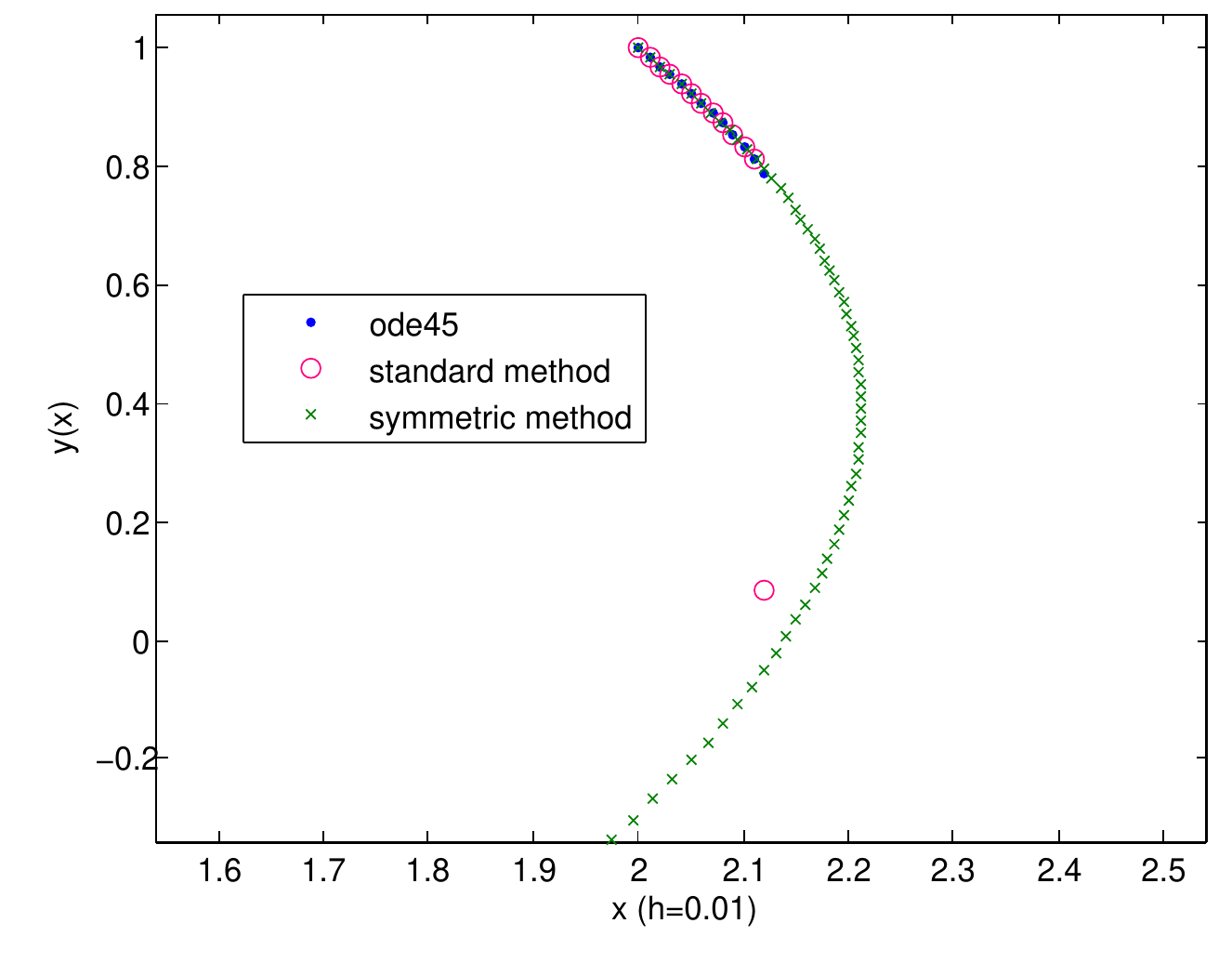}
\caption{All methods for \eref{9} with initial conditions $\{x_0=2,y_0=1,y'_0=-1.5,y''_0=-1.5\}$}
\label{fig4}
\end{center}
\end{figure}

Similarly as for the $sl_3$ realization, the standard method for the 3rd order equation leads to a complicated nonlinear equation while the symmetric scheme is given by
\begin{eqnarray}\label{sl4explicitscheme}
(-2(x_{n+1}-x_{n})+4(x_{n+1}p-x_{n}q_n))x_{n+2}+2(y_{n+1}-y_{n})y_{n+2}\nonumber\\
=y_{n+1}^2-y_{n}^2-(x_{n+1}^2-x_{n}^2)\\
(y_{n+2}-y_{n+1})^2-(x_{n+2}-x_{n+1})^2=4x_{n+2}x_{n+1}p\nonumber
\end{eqnarray}
where $q_n$ and $p$ are constants at each step. Solving the symmetric scheme then amounts to finding the intersection between a straight line and a hyperbola at each step. There is again a geometrical similarity between the discrete scheme and the exact solution for the 2nd order ODE.\\

\section{Conclusions}

From this article and two previous ones \cite{7,8} we conclude the following. For all 4 inequivalent realizations of $sl(2,\mathbb{R})$ symmetry preserving discretization provide qualitatively better numerical solutions than common finite difference methods (including Matlab's solvers), particularly for solutions with singularities. The symmetry preserving methods also provide solutions at a lower computational cost.\\

We would like to emphasize that we are not simply presenting examples when the symmetry preserving numerical methods work. All second order ODEs allowing a nontrivial symmetry algebra were discretized in this manner in \cite{5}. In \cite{6} it was shown that if the symmetries are Lagrangian ones, then the obtained difference systems can be solved exactly (analytically). Article \cite{7} was devoted to second and third order ODEs with three dimensional Lie algebras (solvable, or simple) and mainly discussed the improved precision of symmetry respecting discretizations (at no additional computational cost). The emphasis in \cite{8} and the present article is on the behaviour of solutions at singularities (for all 4 distinct realizations of $sl(2,\mathbb{R})$).

\ack

We thank Anne Bourlioux for helpful discussions. The research of P.W. was partly supported by a grant from NSERC of Canada. R.R. acknowledges a doctoral fellowship from FQRNT.

\end{document}